\documentclass{tlp}

\usepackage{amsmath}
\usepackage{amssymb}
\usepackage{xspace}
\usepackage[inline]{enumitem}
\usepackage{fancyvrb}
\usepackage{xcolor}
\usepackage{soul}
\usepackage{tikz}
\usetikzlibrary{positioning,patterns,shapes,arrows,shadows,backgrounds,fit}
\usepackage[colorlinks,linkcolor=blue,citecolor=blue,urlcolor=blue]{hyperref}

\VerbatimFootnotes

\newcommand{\setlog}{$\{log\}$\xspace}
\newcommand{\agda}{\textsc{Agda}\xspace}
\newcommand{\dafny}{\textsc{Dafny}\xspace}
\newcommand{\whyt}{\textsc{Why3}\xspace}

\newcommand{\defs}{\triangleq}
\DeclareMathSymbol{\dres}{\mathbin}{AMSa}{"43}
\def	\lover		{\mathbin{{\dres} \llap{+\!\!}\;}}
\newcommand{\num}{\mathbb{Z}}

\newcommand{\why}[1]{\tag*{{\footnotesize [by #1]}}}
\newcommand{\Let}[3]{%
  \textsc{let~}#1
  \textsc{~be~}#2
  \textsc{~in~}#3}

\newtheorem{remark}{Remark}

\newif\ifcomments
\commentstrue 

\begin{document}

\lefttitle{Cristi\'a, Capozucca and Rossi}

\jnlPage{1}{8}
\jnlDoiYr{2021}
\doival{10.1017/xxxxx}

\title[\setlog: From a CLP Language to a Formal Verification Tool]
{\setlog: From a Constraint Logic Programming Language to a Formal Verification Tool}

\begin{authgrp}
\author{\sn{Maximiliano} \gn{Cristi\'a}}
\affiliation{Universidad Nacional de Rosario and CIFASIS \\\email{cristia@cifasis-conicet.gov.ar}}
\author{\sn{Alfredo} \gn{Capozucca}}
\affiliation{Universit\'e du Luxembourg \\ \email{alfredo.capozucca@uni.lu}}
\author{\sn{Gianfranco} \gn{Rossi}}
\affiliation{Universit\`a di Parma}
\end{authgrp}

\history{\sub{xx xx xxxx;} \rev{xx xx xxxx;} \acc{xx xx xxxx}}

\maketitle

\begin{abstract}
\setlog (read `setlog') was born as a Constraint Logic Programming (CLP) language where sets and binary relations are first-class citizens, thus fostering set programming. Internally, \setlog is a constraint satisfiability solver implementing decision procedures for several fragments of set theory. Hence, \setlog can be used as a declarative, set, logic programming language and as an automated theorem prover for set theory.
Over time \setlog has been extended with some components integrated to the satisfiability solver thus providing a formal verification environment. In this paper we make a comprehensive presentation of this environment which includes a language for the description of state machines based on set theory, an interactive environment for the execution of functional scenarios over state machines, a generator of verification conditions for state machines, automated verification of state machines, and test case generation. State machines are both, programs and specifications; exactly the same code works as a program and as its specification. In this way, with a few additions, a CLP language turned into a seamlessly integrated programming and automated proof system.
\end{abstract}


\section{Introduction}

The work presented in this paper is aligned with the following quotations \citep{DBLP:conf/fmics/GaravelBP20}.

\begin{quote}
{\small The introduction of programming languages supported by proof systems. I
do expect it to become more common with programming language implementations being
born with program verifiers of various kinds. [\dots] It is a pleasure to see modern programming languages to an increasing degree
look and feel like well known formal specification languages.}

\flushright{\emph{Klaus Havelund}}
\end{quote}

\begin{quote}
{\small Another challenge is to better integrate specification and proof with programming, preferably at the level
of programming languages and tools.}

\flushright{\emph{Xavier Leroy}}
\end{quote}

That is, this paper presents \setlog (read `setlog') \citep{setlog,Dovier00}, a programming language that has the look-and-feel of a formal specification language which in turn is supported by a proof system. The core of \setlog is a Constraint Logic Programming (CLP) language and interpreter written in SWI-Prolog implementing a constraint satisfiability solver for set theory. The satisfiability solver is a rewriting system implementing decision procedures for several fragments of set theory (e.g., integer intervals \citep{DBLP:journals/tocl/CristiaR24} and restricted quantifiers \citep{DBLP:journals/jar/CristiaR24}). In this way, \setlog implements declarative programming in the form of set programming \cite[]{DBLP:books/daglib/0067831,DBLP:series/mcs/CantoneOP01}. As such \setlog has been applied to several case studies showing that the satisfiability solver works in practice \citep{DBLP:journals/jar/CristiaR21,DBLP:journals/jar/CristiaR21b,DBLP:conf/birthday/CristiaR24,DBLP:journals/jar/CristiaLL23,DBLP:conf/csfw/CapozuccaCHK24}.

However, writing good \setlog code, adding the corresponding verification conditions and passing them to the satisfiability solver as to ensure the code verifies some properties, can be cumbersome. In this context, good \setlog code means code for which it is easy to write down verification conditions that can be automatically discharged by \setlog. As a programming language, \setlog includes many programming facilities leading to code that is not always easy to verify because, many times, lies outside the fragments of set theory for which \setlog implements decision procedures. Hence, if a verification condition is given by a \setlog formula for which no decision procedure can be applied, \setlog will not be able to automatically discharge it. This, in turn, implies less reliable code that needs more human effort to be verified.

For these reasons, over the years, \setlog has been extended with some components (also implemented in SWI-Prolog) that constrain programmers to write code that is easier to verify. This programming style within \setlog defines a language for the description of state machines in terms of first-order logic (FOL), set theory and integer arithmetic---much as in formal notations such as B \citep{Abrial00} and Z \citep{DBLP:books/daglib/0072139}. In this way:
\begin{enumerate*}[label=(\emph{\roman*})]
\item the state machine language allows for the automatic generation of verification conditions (VC);
\item these VC are automatically fed into the satisfiability solver which informs the user if they could be discharged or not (in which case, a counterexample is generated);
\item state machines can still be executed as regular \setlog code thus allowing users to use them as working functional prototypes before attempting any serious proof campaign; and
\item users can generate test cases to test a future, more efficient implementation written in some other programming language. 
\end{enumerate*}
This paper makes a comprehensive presentation of these extensions aiming at showing how theoretical contributions in the fields of CLP and set programming can be turned into a feasible tool aimed at formal verification.

As the above quotations suggest, there are other systems where programming and proof coexist. Four conspicuous, widely known, successful representatives of this kind are \agda \citep{DBLP:conf/afp/Norell08,agda}, \dafny \citep{DBLP:conf/lpar/Leino10,dafny}, F* (read `f-star') \citep{DBLP:conf/popl/SwamyHKRDFBFSKZ16,fstar} and \whyt \citep{Bobot00,why3}. All of them have been applied to a range of problems, some times of industrial size and complexity. 

\agda is a typed functional programming language and proof assistant based on intuitionistic type theory. The functional language provides usual functional features such as inductive types, pattern matching, and so on. Proofs are interactive and written in a functional programming style that, unlike Coq (now renamed as Rocq) \citep{DBLP:series/txtcs/BertotC04}, does not provide proof tactics. Programmers can, nonetheless, implement \agda functions that return a proof of some property\footnote{Writing such proofs requires an unusual degree of training in formal software development.}. These functions embody sort of proof tactics. Later, these functions are run during type checking: if one of them fails, then type checking fails. This means that the program does not verify a certain property. As a form of proof automation \agda provides proof search by enumerating possible proof terms that may fit in the specification. 

\dafny combines ideas from the functional and imperative paradigms,  including some support for object-oriented programming. Programs are annotated with specifications through preconditions, postconditions, loop invariants, etc. The specification language has the look-and-feel of a programming language including mathematical integers, bit-vectors, sequences, sets, induction, and so on.  Proof obligations are automatically discharged, given sufficient specification, by the Z3 SMT solver \citep{Bjorner00,Z3}. Sufficient specification means, in general, to write sufficient proof steps (e.g., the inductive step in an inductive proof) the remainder of which are discharged automatically.

F* supports purely functional programming as well as effectful programming \citep{DBLP:journals/jfp/McbrideP08}. Its powerful (dependent) type system allows to describe precise program specifications including security and functional properties. The typechecker will check whether or not the program verifies its specification by discharging proofs interactively written by the programmer (using tactics, metaprogramming and symbolic computation) or by calling the Z3 SMT solver. 

Why3 is a platform for deductive program verification. It provides WhyML, a specification and programming language. Why3 uses many external theorem provers (e.g., Alt-Ergo \citep{Conchon2013,altergo}, CVC4-5
\citep{DBLP:conf/cav/BarrettCDHJKRT11,DBLP:conf/tacas/BarbosaBBKLMMMN22}
and E-prover \citep{eprover,Schulz2019}) to discharge proof obligations. The system includes a library of mathematical theories (e.g., arithmetic and sets) and some programming data structures (e.g., arrays and hash tables). 

Having a different origin than the systems mentioned above and not  being strictly programming and proof systems, we should also mention \textsc{Atelier-B} \citep{DBLP:conf/rssrail/2016,atelierb}, \textsc{Rodin} \citep{DBLP:journals/sttt/AbrialBHHMV10,rodin} and Alloy \citep{DBLP:journals/cacm/Jackson19}.
\textsc{Atelier-B} and \textsc{Rodin} are meant to be used to write and verify B and Event-B \citep{Abrial:2010:MES:1855020} specifications. In both languages, specifications take the form of state machines described by means of FOL and set theory. This is the main point of contact with \setlog. Both tools rely on automatic and interactive provers, including internal ones and external SMT solvers such as VeriT \citep{DBLP:conf/cade/BoutonODF09,veriT} and CVC4. B and Event-B specifications can be executed (or animated) if the \textsc{ProB} tool \citep{DBLP:journals/sttt/LeuschelB08,probweb} is available in \textsc{Atelier-B} and \textsc{Rodin}. \textsc{ProB} is a constraint solver developed in Prolog. Unlike \setlog, in general, \textsc{ProB} cannot prove properties although it can disprove them. 

While earlier versions of Alloy focused mainly on describing and analyzing software structures using FOL and set theory, Alloy 6 also natively supports behavioral modeling.
It provides a unified language for both specifying the system and its expected properties---via linear temporal logic. 
A key difference is that Alloy 6 achieves completeness with respect to trace size, but remains bounded in relation to the cardinality of sets, a limitation that does not apply to \setlog.


\paragraph{Contributions and novelty.}
The main contribution of this paper is a comprehensive presentation of the following \setlog extensions:
\begin{enumerate}
\item A declarative state machine specification language defined on top of the \setlog constraint language. State machines are specified in terms of set theory and FOL.
\item The \textsc{Next} environment and other facilities to simplify the execution and analysis of functional scenarios based on state machines.
\item A verification condition generator (VCG) that generates a set of standard VC for state machines. VC are discharged by calling \setlog itself, and if a VC could not be discharged, it allows users to analyze why this happened.
\item An implementation of a model-based testing framework for \setlog state machines that generates test cases by calling \setlog itself. Users can generate test cases from the state machine to test its implementation.
\end{enumerate}

None of these extensions have been published before nor presented as an integrated verification framework with \setlog at its center.
Although state machines can be specified in \setlog without the language presented here \citep{DBLP:journals/jar/CristiaR21,DBLP:journals/jar/CristiaR21b,DBLP:journals/jar/CristiaLL23}, in that case, verification conditions cannot be automatically generated and the \textsc{Next} environment and automated test case generation are harder to implement, if possible.
Hence, the state machine language is the enabler for the other features listed above.
All the new features are implemented with +3 KLOC or +129 Kb of  SWI-Prolog code distributed across two main files (\verb+setlog_vcg.pl+ and \verb+setlog_ttf.pl+).%
\footnote{The \setlog distribution comprises some other files making the core of the solver.}

Concerning novelty, \setlog is quite different from all the systems mentioned above:
\begin{enumerate*}[label=(\emph{\alph*})]
\item it provides a single language for properties, specifications and programs, there is no distinction between programs and specifications, the same code works as a program and its specification;
\item the same solver is used to execute programs, to prove their properties and to generate test cases, no external solvers are needed;
\item sets and binary relations are first-class entities thus naturally raising the abstraction level of specifications; and
\item all proofs are automatic as the result of implementing decision procedures for several expressive fragments of set theory.
\end{enumerate*}

Although each of these characteristics are not new in the CLP community, as far as we know, no other tool presents all of them together in the context of formal verification. We believe \setlog provides evidence that CLP techniques are as valuable for formal verification as techniques coming from the functional programming realm.

\paragraph{Structure of the paper.}
The paper starts by introducing \setlog in a user-friendly manner in Section \ref{setlog}. 
Section \ref{statemachines} describes the state machine specification language by means of a running example. 
These state machines can be used as functional prototypes. We show how to run and analyze functional scenarios over these prototypes in Section \ref{execution}.
The VCG is described in Section \ref{consistency}. This section includes how \setlog can be used as an automated theorem prover, how to analyze undischarged verification conditions, and an informal account of the classes of formulas fitting in the decision procedures implemented in \setlog. 
Section \ref{testing} describes the model-based testing method implemented in \setlog. 
We close the paper with some concluding remarks in Section \ref{conclusions}.

\RecustomVerbatimEnvironment{Verbatim}{Verbatim}{xleftmargin=8mm}

\section{\label{setlog}The \setlog CLP language and satisfiability solver}

\setlog is a publicly available constraint satisfiability solver and a declarative set-based, constraint-based programming language implemented in SWI-Prolog \citep{setlog,Dovier00}.
\setlog is deeply rooted in the work on Computable Set Theory \citep{10.5555/92143}, combined with the ideas put forward by the set-based programming language SETL \citep{DBLP:books/daglib/0067831}.

\setlog implements various decision procedures for different theories on the domain of finite sets and integer numbers. Specifically, \setlog implements decision procedures for:
\begin{enumerate*}[label=(\arabic*)]
\item\label{i:set} The theory of \emph{hereditarily finite sets}, i.e., finitely nested sets that are finite at each level of nesting \citep{Dovier00};
\item A very expressive fragment of the theory of finite set relation algebras \citep{DBLP:journals/jar/CristiaR20,DBLP:conf/RelMiCS/CristiaR18}; 
\item Theory \ref{i:set} extended with restricted intensional sets \citep{DBLP:journals/jar/CristiaR21a};
\item\label{i:card} Theory \ref{i:set} extended with cardinality constraints \citep{DBLP:journals/tplp/CristiaR23}; 
\item Theory \ref{i:card} extended with integer intervals \citep{DBLP:journals/tocl/CristiaR24};
\item Quantifier-free, decidable languages extended with restricted quantifiers \citep{DBLP:journals/jar/CristiaR24}; and
\item The theory of linear integer arithmetic, by integrating an existing decision procedure for this theory.
\end{enumerate*}
All these
procedures constitute the core of the \setlog tool. Several in-depth empirical
evaluations provide evidence that \setlog is able to solve non-trivial problems
\citep{DBLP:journals/jar/CristiaR20,DBLP:conf/RelMiCS/CristiaR18,DBLP:journals/jar/CristiaR21a,CristiaRossiSEFM13};
in particular as an automated verifier of security properties
\citep{DBLP:journals/jar/CristiaR21,DBLP:journals/jar/CristiaR21b,DBLP:journals/jar/CristiaLL23,DBLP:conf/csfw/CapozuccaCHK24}.

\begin{remark}[Scope of presentation]
Below we introduce \setlog. The presentation is brief and user-oriented. A complete, user-oriented presentation can be found in the \setlog user's manual \citep{Rossi00}. A formal presentation can be found in the papers cited in the previous paragraph, including formal syntax and semantics; and soundness, completeness and termination proofs. We assume a basic knowledge of (constraint) logic programming and Prolog.
\end{remark}

In the introduction we say that in \setlog there is no distinction between programs and specifications. Let's say we need a program computing the minimum element of a set of integer numbers. The specification of that program can be the following predicate where $S$ is a set and $m$ its minimum element ($\defs$ means equal by definition):

\begin{equation}\label{eq:smin}
smin(S,m) \defs m \in S \land (\forall x \in S: m \leq x)
\end{equation}
The \setlog implementation of $smin$ is the following predicate:
\refstepcounter{equation}\label{eq:smini}
\begin{Verbatim}[commandchars=\\\{\}]
smin(S,M) :- M in S & foreach(X in S, M =< X).\hfill\textrm{(\arabic{equation})}
\end{Verbatim}
Right at first glance the differences between specification and program look minimal. For instance, we write \verb+M+ instead of $m$ to denote a variable (as in Prolog), \verb+in+ denotes set membership ($\in$), \verb+&+ denotes conjunction ($\land$), and \verb+foreach+ denotes the restricted universal quantifier ($\forall x \in S:\dots$). Indeed, the following is another implementation for $smin$:
\begin{Verbatim}
smin1(S,M) :- foreach(X in S, M =< X) & M in S.
\end{Verbatim}
That is, \verb+&+ enjoys commutativity as much as $\land$. In other words, both \verb+smin+ and \verb+smin1+ will produce the same results---although they can have different performance. The commutativity of \verb+&+ makes \verb+smin+ look as a formula. But it is also a program as we can call it to get the minimum element of a set:
\begin{Verbatim}
{log}=> smin({9,-3,5,1,8},M).
  M = -3
\end{Verbatim}
Hence, \verb+smin+ is called with an input set and \setlog computes its minimum (\verb+M = -3+). Furthermore, the input set can be partially specified (i.e., some elements can be variables)\footnote{The computed answers are simplified for presentation purposes.}:
\begin{Verbatim}
{log}=> smin({9,-3,5,X,8},M).    % variable X in the set
  M = -3
  Constraint: -3 =< X
  ;
  M = X
  Constraint: X =< 9, X =< -3, X =< 5, X =< 8
\end{Verbatim}
In \setlog a solution is a (possibly empty) list of equalities of the form $var = term$ and a (possibly empty) list of constraints. The list of constraints is guaranteed to be always satisfiable. 
In fact we can see a possible value for \verb+X+ if we ask \setlog to produce \emph{ground solutions} (i.e., solutions where variables occur only at the left-hand side of bindings):
\begin{Verbatim}
{log}=> groundsol.
{log}=> smin({9,-3,5,X,8},M).
  X = -3,  M = -3
\end{Verbatim}

As in Prolog, \setlog does not distinguish between inputs and outputs. The fact that so far we have considered that \verb+S+ is an input and \verb+M+ is an output is purely out of convenience and common use. \setlog can compute the other way round:
\begin{Verbatim}[numbers=left]
{log}=> smin(S,10).
  S = {10 / _N1}
  Constraint: set(_N1), foreach(_X1 in _N1, 10 =< _X1)
{log}=> groundsol.
{log}=> smin(S,10).
  S = {10}
\end{Verbatim}
In the first run (line 1), \setlog states that for 10 to be the minimum of \verb+S+, then \verb+S+ must be a set of the form \verb+{10/_N1}+ meaning that 10 belongs to \verb+S+ and that all of its other possible elements (represented by set \verb+_N1+) are greater than or equal to 10. Again, the constraint is satisfiable---in this case by substituting \verb+_N1+ by the empty set, which is the solution computed after the activation of \verb+groundsol+ (lines 4-6). The term \verb+{10/_N1}+ corresponds to $\{10\} \cup \_N1$ what implies that \verb+{10/{}}+ equals \verb+{10}+.

\verb+smin+ is not a formula only because of the commutativity of \verb+&+ but, fundamentally, because it is possible to prove properties true of it. The following is a known property of the minimum of a set:
\begin{equation}
smin(A,m) \land smin(B,n) \land A \subseteq B \implies n \leq m
\end{equation}
Given that \setlog is a satisfiability solver, in order to check whether or not \verb+smin+ verifies this property we have to ask \setlog if the negation of the property is unsatisfiable. So we run:
\begin{Verbatim}
{log}=> neg(smin(A,M) & smin(B,N) & subset(A,B) implies N =< M).
  no
\end{Verbatim}
As \setlog answers \verb+no+ we know the formula is unsatisfiable---i.e., there are no finite sets \verb+A+ and \verb+B+ and integers \verb+M+ and \verb+N+ that can satisfy the formula. Therefore, the formula inside \verb+neg+ is always satisfiable or valid.

Hence, we can use \verb+smin+ as both a formula or specification and as a program or implementation. The same piece of code, no need for text describing the specification and another text describing the implementation, no need for program annotations. 
We call this the \emph{program-formula duality} or the \emph{implementation-specification duality}. 
Some times we refer to \setlog code as \emph{forgrams}, a portmanteau word resulting from the combination of \emph{for}mula and pro\emph{gram}.
Therefore, when \setlog programmers write code they are writing a
\emph{program as a formula} \citep{Apt1999}.
The same solver executes a program and proves properties true of it, no need for external solvers or theorem provers, at least for its decidable fragments.
Actually, \setlog cannot be divided into a component running programs and a component proving properties.

Although some of these ideas are not new to the CLP community, it is our understanding that no other tool features set programming based on decision procedures implemented with CLP properties---e.g., the program-formula duality.

\subsection{\label{moresets}Extensional sets, relations and functions}
The term $\{x/A\}$ is called extensional set; the empty set is denoted by $\{\}$. These are the basic constructors for sets; they can be arguments to all classic set operators---union, intersection, cardinality, etc. Set operators are provided as constraints. For example, \verb+un(A,B,C)+ denotes $C = A \cup B$, \verb+inters(A,B,C)+ denotes $C = A \cap B$, and \verb+diff(A,B,C)+ denotes $C = A \setminus B$. Then, we can compute the union of two sets even when they are partially specified\footnote{For brevity, only the first solution is shown.}:
\begin{Verbatim}
{log}=> un({X/A},{3,a/B},{Y/C}).
  Y = X,  C = {3,a/_N1}
  Constraint: X nin A, un(A,B,_N1), X neq 3, X neq a, X nin _N1
\end{Verbatim}
where \verb+X nin A+ denotes $X \notin A$ and \verb+X neq 3+ denotes $X \neq 3$---note that \verb+a+ is a constant, not a variable.

The constraint \verb+size(A,N)+ denotes $|A| = N$, i.e., set cardinality. The first argument can be an extensional set or the empty set; the second one can only be a variable or an integer number, but it can participate in integer constraints. Hence, \setlog can compute the solutions (or lack thereof) of\footnote{\Verb+is+ is the standard Prolog predicate for the evaluation of integer expressions.}:
\begin{Verbatim}
{log}=> groundsol.
{log}=> size(A,N) & size(B,M) & B neq {} & 5 is M + N & 2 =< N - M.
  A = {n0,n1,n2,n3},  N = 4,  B = {n4},  M = 1
\end{Verbatim}

In \setlog a binary relation is just a set of ordered pairs. The ordered pair $(x,y)$ is written \verb+[x,y]+. \setlog provides all the operators of set relation algebra in the form of constraints. Then, for example, \verb+comp(R,S,T)+ denotes composition of binary relations ($T = R; S$) and \verb+inv(R,S)+ denotes the converse (or inverse) of a binary relation $(S = R^\smile$). Cartesian product is also available: \verb+cp(A,B)+ denotes the set $A \times B$. Extensional sets and the empty set can be arguments to all relational operators and Cartesian product.

By combining set and relational operators it is possible to define a number of other useful operators. In this way, \setlog provides constraints for the domain and range of binary relations, domain and range restriction, relational image and the widely used `update', `overriding' or `oplus' operator---denoted $\oplus$ in Z and $\lover$ in B. 

Functions are a subclass of binary relations---i.e., functions are sets of ordered pairs. The constraint \verb+pfun(F)+ states that the binary relation \verb+F+ is a function. \setlog provides two notions of function application: if \verb+pfun(F)+ holds then \verb+apply(F,X,Y)+ is satisfied iff $X \in dom(F) \land F(X) = Y$; in turn, \verb+applyTo(F,X,Y)+ denotes $\{(X,X)\}; F = \{(X,Y)\}$, for any binary relation $F$.
That is, \verb+applyTo(F,X,Y)+ is true if \verb+F+ is a binary relation containing exactly one pair whose first
component is \verb+X+ and whose second component is \verb+Y+---meaning that \verb+F+ is not necessarily a function.

Given that relations and functions are sets, they can be combined and passed as arguments to all the classic set and relational operators.

\subsection{\label{setterms}Integer intervals and restricted intensional sets}

Two more set terms are available in \setlog. Terms of the form \verb+int(m,n)+, with both arguments either variables or integer numbers, denote integer intervals---i.e., sets of the form $[m,n] \cap \num$. Interval limits can participate in integer constraints. Intervals can be arguments to all the classic set operators including cardinality---but in general they cannot be passed as arguments to relational operators. In this way \setlog is able to compute solutions (or lack thereof) for goals such as:
\begin{Verbatim}
{log}=> groundsol.
{log}=> un({X,Y},{Z/A},int(5,M)) & X neq Y & 7 < M.
  X = 5,  Y = 6,  Z = 7,  A = {8},  M = 8
\end{Verbatim}

Finally, \setlog provides a set term, called Restricted Intensional Set (RIS), denoting set comprehensions of the form $\{x \in A: \phi(x)\}$ where $\phi$ is some formula depending on $x$. For instance, the set $\{x \in A \mid 3x + 2 \geq 0\}$ is encoded in \setlog as the term \verb.ris(X in A, 3*X + 2 >= 0)., where \verb+A+ can be an extensional set term or another RIS. RIS have a more expressive and complex structure that we are not going to show here. RIS terms can only be used as arguments of classic set operators not including cardinality---they cannot be passed as arguments to relational operators.

\subsection{Negation}

Negation in \setlog is a delicate matter as it is in logic programming in general \citep{DBLP:journals/jlp/AptB94}. The problem with negating a \setlog formula is that its result may not be a \setlog formula. Consider the following alternative implementation of $smin$ \eqref{eq:smin}:
\begin{Verbatim}
smin2(S,M) :- M in S & subset(S,int(M,Max)).
\end{Verbatim}
Note the presence of \verb+Max+, an unbound variable making part of the body but not of the predicate's head. This is an \emph{existential variable}. Thus, the negation of \verb+smin2+ requires the introduction of a (unrestricted) universal quantification over \verb+Max+, that is:
\begin{equation}\label{eq:nsmin2}
\forall Max(\lnot(M \in S \land S \subseteq [M,Max])).
\end{equation}
The problem is that \setlog does not admit (unrestricted) universal quantification, meaning that \eqref{eq:nsmin2}, as it is, cannot be encoded as a \setlog formula. That does not mean that there is no \setlog formula encoding the negation of \verb+smin2+. Actually, the following is such a formula:
\begin{equation}
M \notin S \lor X \in S \land X < M
\end{equation}
were $X$ is an existential variable. The point is that, in general, \verb+neg(smin2)+ cannot be automatically computed. Fortunately, in the case of $smin$, since its encoding with \verb+smin+ in \eqref{eq:smini} contains no existential variables, \setlog is able to automatically compute \verb+neg(smin)+:
\begin{equation}
\lnot(M \in S \land \forall X \in S: M \leq X) \equiv M \notin S \lor \exists X \in S: X < M
\end{equation}
where the right-hand side predicate is encoded in \setlog as:
\begin{Verbatim}
M nin S or exists(X in S, X < M).
\end{Verbatim}

Now consider the following \setlog predicate:
\begin{Verbatim}
property(R,S) :- dom(R,Dr) & dom(S,Ds) & subset(Dr,Ds).
\end{Verbatim}
where \verb+dom(R,D)+ encodes the domain of relation \verb+R+, i.e., $dom(R) = D$. Clearly, \verb+property+ introduces two existential variables (\verb+Dr+ and \verb+Ds+) thus making it impossible to use \verb+neg+ to compute its negation. However, the nature of these variables is different from \verb+Max+ in the $smin$ example. Here, these variables \emph{define} the domain of \verb+R+ and \verb+S+. In other words, \verb+Dr+ and \verb+Ds+ are sort of names for the expressions $dom(R)$ and $dom(S)$. This means that the negation of \verb+property+ is not:
\begin{equation}
\forall Dr,Ds(dom(R) \neq Dr \lor dom(S) \neq Ds \lor Dr \not\subseteq Ds)
\end{equation}
It makes no sense to state $dom(R) \neq Dr$ for all $Dr$ because it would mean that the domain of $R$ does not exist. For cases like this \setlog provides the \textsc{Let} predicate \cite[Sect. 5]{DBLP:journals/jar/CristiaR24}:
\begin{Verbatim}
property(R,S) :- let([Dr,Ds], dom(R,Dr) & dom(S,Ds), subset(Dr,Ds)).
\end{Verbatim}
\textsc{Let} is interpreted as follows:
\begin{align*}
&\Let{x_1,\dots,x_n}{x_1 = e_1,\dots,x_n = e_n}{\phi} \\
&\defs \exists x_1,\dots,x_n (x_1 = e_1 \land \dots \land x_n = e_n \land \phi) \\
& \equiv \phi[e_1/x_1,\dots,e_n/x_n]
\why{one-point rule \cite[Sect 4.2]{DBLP:books/daglib/0072139}}
\end{align*}
where $x_i$ are variables, $e_i$ terms such that $x_i$ does not occur in $e_i$ and $\phi$ is a formula. 
Therefore, negating \verb+property(R,S)+ proceeds as follows:
\begin{align*}
&\lnot property(R,S) \\
&\equiv \lnot(\Let{Dr,Ds}{Dr = dom(R), Ds = dom(S)}{Dr \subseteq Ds})\\
&\equiv \lnot dom(R) \subseteq dom(S)
\end{align*}
where the rightmost formula can be encoded in \setlog as follows:
\begin{Verbatim}
dom(R,Dr) & dom(S,Ds) & neg(subset(Dr,Ds))
\end{Verbatim}

The \textsc{Let} predicate allows us to use \verb+neg+ for a whole class of formulas for which, otherwise, negation would have needed human intervention. In summary, \verb+neg(+$P$\verb+)+ is safe provided $P$ does not contain existential variables.

\subsection{\label{rq}Restricted quantifiers}

We have already introduced restricted quantifiers (RQ) in \setlog. We have seen \verb+foreach+ (restricted universal quantifier, RUQ) and \verb+exists+ (restricted existential quantifier, REQ), in the previous sections. Here we discuss them a little more deeply. 
First of all, RQ in \setlog are at least as expressive as set relation algebra \cite[Sect. 4.3]{DBLP:journals/jar/CristiaR24}, in part, due to the following characteristics.
First, RQ admit \emph{quantification terms} rather than simply variables. For example:
\begin{Verbatim}
foreach([X,Y] in R, X is Y + 1)
\end{Verbatim}
states that all the elements of \verb+R+ of the form \verb+[X,Y]+ must verify \verb!X is Y + 1!. As \verb+[X,Y]+ denotes an ordered pair, then \verb+R+ is expected to be a binary relation.

Second, RQ can be nested. For instance:
\begin{Verbatim}
foreach([[X,Y] in R, [V,W] in S], X = V implies Y is W + 2)
\end{Verbatim}
is equivalent to the following nested RUQ:
\begin{Verbatim}
foreach([X,Y] in R, foreach([V,W] in S, X = V implies Y is W + 2))
\end{Verbatim}
Third, a bound variable can be used as quantification domain in an inner RQ:
\refstepcounter{equation}\label{eq:rq}
\begin{Verbatim}[commandchars=\\\{\}]
foreach([X,Y] in R, exists(V in Y, X is V + 1)) \hfill\textrm{(\arabic{equation})}
\end{Verbatim}
That is, the elements of the range of \verb+R+ are expected to be sets---due to the set membership constraint \verb+V in Y+.

The introduction of REQ in the language deserves a special note. Since logic programming allows for the introduction of existential variables, the presence of REQ might seem unnecessary, but it is not. REQ help in further extending the fragment of the language where negation can be easily computed. Consider the following simple predicate:
\begin{Verbatim}
p(S) :- X in S & X > 0.
\end{Verbatim}
The negation of \verb+p+ cannot be easily computed due to the presence of \verb+X+, an existential variable. However, \verb+p+ can be written in terms of a REQ preserving its meaning while exchanging an existential variable for a bound one:
\begin{Verbatim}
p(S) :- exists(X in S, X > 0).
\end{Verbatim}
Now \verb+neg(p(S))+ can be easily computed:
\begin{equation}
\lnot p(S) \equiv \lnot (\exists X \in S: X > 0) \equiv (\forall X \in S: \lnot(X > 0))  \equiv (\forall X \in S: X \leq 0)
\end{equation}
where the rightmost predicate is written in \setlog as: \verb+foreach(X in S, X =< 0)+.

\subsection{Types}

Rooted in Prolog, \setlog provides essentially an untyped  language  based on untyped FOL and untyped set theory. In this way, a set such as $\{1,a,(3,b),\{x,q\}\}$ is perfectly legal in \setlog. Untyped formalisms are not bad in themselves but types help to avoid some classes of errors \citep{DBLP:journals/toplas/LamportP99}. For this reason, we recently defined a type system and implemented a typechecker as a \setlog component \citep{DBLP:journals/tplp/CristiaR24}. 

Users can activate/deactivate the typechecker according to their needs. The type system defines basic or uninterpreted types, types for integers and strings, sum types\footnote{Tagged union, variant, variant record, choice type, discriminated union, disjoint union, or coproduct.}, Cartesian products, and set types---in this sense, the type system is similar to those used in B and Z. Set and relational operators are polymorphic. Users must declare the type of user-defined predicates. We will see more about types in Section \ref{statemachines}.

\subsection{Undefinedness}

In B and Z, undefinedness arises whenever a partial function is applied to an argument that does not belong to the domain of the function.
In turn, undefinedness may lead to inconsistencies.
In those notations undefinedness is solved by requesting proof obligations guaranteeing that the argument belongs to the domain of the partial function.

In \setlog undefinedness is less pervasive because function application becomes a predicate rather than a term, as shown in Section \ref{moresets}.
In effect, if \verb+X+ does not belong to the domain of \verb+F+, \verb+apply(F,X,Y)+ is simply unsatisfiable.
In other words, function application cannot remain undefined: it is either satisfiable or unsatisfiable.
Same considerations hold for \verb+applyTo+.

Another situation for undefinedness is division by zero in integer arithmetic.
In this case B and Z requires to prove that $y \neq 0$ whenever $x \mathbin{\mathtt{div}} y$ is in context.
In \setlog this situation does not occur given that it only supports linear integer arithmetic which rules out expressions such as $x \mathbin{\mathtt{div}} y$.
Whenever \setlog tries to solve a predicate including such a term an exception is raised.

In summary, in \setlog function application does not lead to inconsistencies.

\section{\label{statemachines}A logic language for state machines}
As we have explained in the introduction, we have defined a language on top of \setlog for the description of state machines. The language is inspired in the B and Z specification languages.
This language constrains programmers to use \setlog in a more restricted way but increasing the chances of automatic verification.

We will introduce the main elements of the language and the functionalities described in coming sections by means of a running example.
Some elements are deliberately left out of the paper for brevity.
More details can be found in the \setlog user's manual \cite[Section 13]{Rossi00} and in 10 case studies that are publicly available.%
\footnote{Case studies and running example: \href{https://www.clpset.unipr.it/SETLOG/APPLICATIONS/fv.zip}{https://www.clpset.unipr.it/SETLOG/APPLICATIONS/fv.zip}.}
These case studies include informal requirements, the \setlog specification of a state machine, \textsc{Next} scenarios (Section \ref{next}), VCG commands (Section \ref{consistency}) and test case generation (Section \ref{testing}).
These case studies make use of some features not included in the running example such as RUQ, cardinality, integer arithmetic, set relation algebra, etc.

The specification to be used as running example is known as the \emph{birthday book}.
It is a system which records people's birthdays, and is
able to issue a reminder when the day comes round. The problem is borrowed from Spivey \citeyearpar{Spivey00}.

A \setlog state machine is a collection of \emph{declarations} and \emph{predicates}.
Predicates and declarations are similar to Prolog's.
State machines transition between states defined by state variables.
The first declaration is used to declare the state variables of the state machine. In the birthday book we have the following declaration:
\begin{Verbatim}
variables([Known, Birthday]).
\end{Verbatim}
where \verb+Known+ is intended to be the set of names with birthdays recorded; and \verb+Birthday+ is a function which, when applied to certain names, gives the birthdays associated with them.

After introducing the state variables, one or more state invariants can be declared. In the birthday book we have the following two:
\begin{Verbatim}
invariant(domBirthday).
domBirthday(Known,Birthday) :- dom(Birthday,Known).
\end{Verbatim}
\begin{Verbatim}
invariant(birthdayFun).
birthdayFun(Birthday) :- pfun(Birthday).
\end{Verbatim}
As can be seen, each invariant is preceded by an \verb+invariant+ declaration. The body of each invariant is a \setlog formula depending on the state variables.
An invariant declaration is a declaration of intent.
It remains to be proved that these predicates are indeed preserved by the operations of the specification---see Section \ref{consistency}.
By the end of Section \ref{undischarged} we analyze the possibility of conjoining these invariants in just one declaration.

In the state machine language recursion is not allowed---although \setlog admits recursive predicates. Hence, invariants, as well as all the other elements of the language, have to be given as non-recursive \setlog predicates. In spite that this can be seen as a severe restriction, it is not. The main elements of specification languages such as Z and B do not admit recursive definitions. Given the applicability of Z and B, the lack of recursion in the \setlog state machine language should not be a hard limitation.

Once all the invariants have been given, the set of initial states of the specification has to be given. In many specifications a single initial state is given. This is the case of the birthday book.
\begin{Verbatim}
initial(bbInit).
bbInit(Known,Birthday) :- Known = {} & Birthday = {}.
\end{Verbatim}

The main part of a \setlog state machine specification is the definition of the operations (state transitions) of the specification.
Operations depend on state variables and input and output parameters.
Besides, for each state variable $V$ the predicate can also depend on $V\_$, which represents the value of $V$ in the next state\footnote{$V\_$ plays the same role than $V'$ in Z specifications. We chose to use `\_' instead of the prime character due to Prolog compatibility.}. The first operation of the birthday book adds a name and the corresponding birth date to the system.
Hence, the head of the clause is the following:
\begin{Verbatim}
addBirthday(Known,Birthday,Name,Date,Known_,Birthday_)
\end{Verbatim}
where \Verb+Name+ and \Verb+Date+ are the inputs; and \Verb+Known+ and \Verb+Birthday+ represent the before-state while \Verb+Known_+ and \Verb+Birthday_+ represent the after-state.
We will define \verb+addBirthday+ by splitting it in a couple of predicates. The first one specifies the case when the given name and date can actually be added to the system, i.e., when the name is new to the system:
\begin{Verbatim}
addBirthdayOk(Known,Birthday,Name,Date,Known_,Birthday_) :-
  Name nin Known &
  un(Known,{Name},Known_) &
  un(Birthday,{[Name,Date]},Birthday_).
\end{Verbatim}
It is easy to see that the specification is given by providing pre- and postconditions.
For instance, the constraint \Verb+Name nin Known+ is a precondition whereas \verb+un(Birthday,{[Name,Date]},Birthday_)+ is a postcondition stating that the value of \verb+Birthday+ in the next state is equal to the union of its value in the before state with \verb+{[Name,Date]}+.

The second predicate describes what to do if the user wants to add a name already present in the system:
\begin{Verbatim}
nameAlreadyExists(Known,Birthday,Name,Known_,Birthday_) :-
  Name in Known & Known_ = Known &  Birthday_ = Birthday.
\end{Verbatim}
That is, the system has to remain in the same state. Finally, we declare the full operation:
\begin{Verbatim}
operation(addBirthday).
addBirthday(Known,Birthday,Name,Date,Known_,Birthday_) :-
  addBirthdayOk(Known,Birthday,Name,Date,Known_,Birthday_)
  or
  nameAlreadyExists(Known,Birthday,Name,Known_,Birthday_).
\end{Verbatim}
In this case we precede the clause by the \verb+operation+ declaration. Note that \verb+addBirthdayOk+ and \verb+nameAlreadyExists+ are not operations, although they participate in one.

Now we specify an operation listing all the persons whose birthday is a given date:
\begin{Verbatim}
operation(remind).
remind(Birthday,Today,Names) :-
  rres(Birthday,{Today},M) & dom(M,Names).
\end{Verbatim}
where \verb+rres(R,B,S)+ is a constraint called \emph{range restriction} whose interpretation is $S = \{(x,y) \in R \mid y \in B\}$\footnote{\Verb+rres/3+ is one of the relational operators that can be defined in terms of the classic set operators and the basic operators of set relation algebra.}. Note that state variable \verb+Known+ is not an argument simply because the operation does not need it. Here \verb+Today+ is (supposed to be) an input whereas \verb+Names+ is (supposed to be) an output. This distinction is enforced by the user when the specification is executed. \verb+remind+ does not transition to a new state; it just outputs information. This is specified by \emph{not} including the next-state variables in the head of the clause.

One more operation of the birthday book specification is given in  \ref{bb}.

In summary, the specification is given by a declarative, abstract, logic description---similar to those written in B and Z. Sets, binary relations, functions and their operators are the main building blocks of the language. Properties (invariants) and operations are given in the same and only language.

\subsection{Model parameters, axioms and user-defined theorems}

The state machine specification language also supports  \emph{parameters}, \emph{axioms} and user-defined \emph{theorems} \citep[Section 13.1.1]{Rossi00}. 
Parameters play the role of machine parameters and constants in B specifications and the role of variables declared in axiomatic definitions in Z. That is, parameters serve to declare the existence of some (global) values accessible to invariants and operations, but they cannot be changed by operations---i.e., there is no next-state value for a parameter. 
Axioms are used to state properties of parameters. 
User-defined theorems are used to state properties that can be deduced from axioms, invariants, operations or theorems that have already been declared.

\subsection{Typing state machines}
So far we have paid no attention to types in the specification. However, users can, but are not forced to, add typing information to each predicate (invariants, operations, etc.) declared in the specification to avoid some classes of errors. Here we type \verb+remind+ just as an example.
\begin{Verbatim}
operation(remind).
dec_p_type(remind(rel(name,date),date,set(name))). % type declaration
remind(Birthday,Today,Names) :-
  rres(Birthday,{Today},M) & dom(M,Names) & dec(M,rel(name,date)).
\end{Verbatim}
The \verb+dec_p_type+ declaration provides the type for \verb+remind+. The term \verb+remind(rel(name,date),date,set(name))+ declares the type of each argument of the operation. For instance, \verb+rel(name,date)+ is the type of \verb+Birthday+, \verb+date+ is the type of \verb+Today+, and so on. The type of \verb+M+ is given by an explicit type declaration, \verb+dec(M,rel(name,date))+. In turn, \verb+name+ and \verb+date+ are basic or uninterpreted types; \verb+rel(name,date)+ is the type of all binary relations with domain in \verb+name+ and range in \verb+date+; and \verb+set(name)+ is the type of all sets whose elements are of type \verb+name+. If \verb+t+ is a basic type then its elements are of the form \verb+t:+$\langle elem \rangle$, where $elem$ is any Prolog atom or natural number.

\section{\label{execution}Machine execution}

As we have shown in Section \ref{setlog}, \setlog predicates are both, programs and specifications. State machines defined as in the previous section are no exception. Then, the operations of the specification of the birthday book can be executed as if they were the routines of some program. As \setlog programs are normally much less efficient than Prolog programs, we see them as functional prototypes.
Running functional scenarios on a prototype helps, for example, to uncover possible mistakes in the specification early on, to analyze complex features, to analyze interactions among the operations, novice users to check if they have written the right predicates, etc.
Functional scenarios can be executed directly from the \setlog interpreter in two different ways: by using the \textsc{Next} environment (Section \ref{next}), and by using symbolic execution (Section \ref{symbexec}).

\subsection{\label{next}The \textsc{Next} environment}

\textsc{Next} is a component tightly integrated with \setlog that simplifies the execution and analysis of deterministic, fully instantiated functional scenarios.
Although \textsc{Next} is less general than the approach shown in the next section, it considerably eases the job of users. 
For example, we can add a person's name and birth date when the system is in its initial state, as follows:
\begin{Verbatim}
{log}=> initial >> addBirthday(Name:alice,Date:may24).
  Known = {alice}, Birthday = {[alice,may24]}
\end{Verbatim}
\verb+initial+ refers to the predicate declared as such in the specification; \verb+>>+ (read \emph{then}) imposes a sequential order in the execution.
\setlog automatically fetches state variables when calling an operation; users need to indicate values only for the input variables (e.g., \verb+Name:alice+).

With \textsc{Next} it is easy to add more people to the system:
\begin{Verbatim}
{log}=> initial >>
        addBirthday(Name:[[alice,bob]],Date:[[may24,nov05]]).
  Known = {alice,bob},  Birthday = {[alice,may24],[bob,nov05]}
\end{Verbatim}
In fact the above call is equivalent to the following one:
\begin{Verbatim}
{log}=> initial >>
        addBirthday(Name:alice,Date:may24) >>
        addBirthday(Name:bob,Date:nov05).
\end{Verbatim}
As can be seen, \textsc{Next} automatically chains the after-state of an operation with the before-state of the next one thus relieving users from having to introduce new variables to get each state of the system. 

More operations can be called by using \Verb+>>+ to execute more complex scenarios.
Moreover, the execution trace of the scenario can be analyzed by enclosing the initial state between square brackets, as follows:
\begin{Verbatim}
{log}=> [initial] >>
        addBirthday(Name:[[alice,bob]],Date:[[may24,nov05]]) >>
        remind(Today:may24,Cards).
\end{Verbatim}
in which case the full state trace is shown as follows:
\begin{Verbatim}
Known = {},  
Birthday = {}
  ----> addBirthday(Name:alice,Date:may24)
Known = {alice},  Birthday = {[alice,may24]}
  ----> addBirthday(Name:bob,Date:nov05)
Known = {alice,bob},  Birthday = {[alice,may24],[bob,nov05]}
  ----> remind(Today:may24,Cards)
Known = {alice,bob},  Birthday = {[alice,may24],[bob,nov05]},  
Cards = {alice}
\end{Verbatim}
The execution starts from the initial state (\verb+Known = {}, Birthday = {}+), then Alice's birthday is added thus arriving to the state given by \verb+Known = {alice}, Birthday = {[alice,may24]}+.
Afterwards, Bob's birthday is added to the book and finally we call for the set of people whose birthday is May, 24.
In this case the answer is \verb+Cards = {alice}+ and we can also check that \verb+remind+ does not change the state of the system.

Whenever \textsc{Next} cannot fully determine the next state or an output value, an error message is issued.
\begin{Verbatim}
{log}=> initial >> addBirthday(Name,Date:may24).
ERROR: (next) next-state/output not sufficiently instantiated after:
       addBirthday(Name,Date:may24)
\end{Verbatim}
Since there is no value bound to \verb+Name+ the next state of the system remains underspecified making \textsc{Next} to issue an error.
This helps users to analyze the behavior of the system in situations similar to the real usage.

\textsc{Next} also performs invariant checking after each step of an execution. Users can indicate the invariants to be checked by appending a list with their names to the initial state. For example: \verb+[initial]:[domBirthday]+ would make \textsc{Next} to check \Verb+domBirthday+ after each step of a scenario. If an invariant is not satisfied after a given step, then \textsc{Next} informs the user of the situation. This, in turn, provides valuable information on the correctness of the specification.

\subsection{\label{symbexec}Symbolic executions}

\setlog can perform symbolic executions too \citep{DBLP:journals/cacm/King76}. That is, we can execute the prototype by providing some variables instead of constants as inputs---or a mixture of both. In this way we can draw more general conclusions about the specification. In this case, though, we cannot use \textsc{Next} because it requires variables involved in the scenario to be completely instantiated. Actually, in symbolic executions we need to analyze the constraints returned by \setlog. The following scenario analyses how \verb+remind+ responds to a call where \verb+Birthday+ has at least two equal dates in its range:
\begin{Verbatim}
{log}=> pfun({[X,Y],[Z,Y]/B}) & remind({[X,Y],[Z,Y]/B},Y,Names).
  Names = {X,Z/_N1}
  Constraint: X neq Z, dom(_N5,_N1), un(_N4,_N5,B), ran(_N5,_N3),
              subset(_N3,{Y}), ...
\end{Verbatim}
We have deliberately cut out the list of constraints returned by \setlog to simplify the exposition.
\verb+pfun({[X,Y],[Z,Y]/B})+ restricts the set to be a function, thus forcing \verb+X neq Z+.
Then, the names returned by \verb+remind+ include \verb+X+ and \verb+Z+, plus some set \verb+_N1+.
The constraints state that \verb+_N1+ is the domain of \verb+_N5+, a subset of \verb+B+, whose range contains only \verb+Y+.
That is, \verb+Names+ contains the right names and nothing else.

Clearly, this kind of answers is harder to analyze than the answers given by \textsc{Next} although the former yield more general results.
To help during the symbolic analysis of the system users can combine it with ground solutions.
For example, the answer to the above scenario when executed in \verb+groundsol+ mode (Section \ref{setlog}) is the following:
\begin{Verbatim}
X = n0,  Y = n2,  Z = n1,  B = {},  Names = {n0,n1}
\end{Verbatim}
We can see that \verb+Names+ contains just the values of \verb+X+ and \verb+Z+, thus coinciding with the above analysis.

Another scenario that can be analyzed is given by \verb+Birthday+ being a constant function:
\begin{Verbatim}
{log}=> pfun({[X,Y],[Z,Y]/B}) & ran(B,R) & subset(R,{Y}) & 
        remind({[X,Y],[Z,Y]/B},W,Names).
\end{Verbatim}
Note that in this case \verb+W+, not \verb+Y+, is the second argument passed in to \verb+remind+.
We will analyze only two solutions:
\begin{enumerate*}[label=(\emph{\roman*})]
\item Obtaining the reminders for date \verb+Y+, which ensures that at least \verb+X+ and \verb+Z+ are contained in the list of names; and
\item Obtaining reminders for any other date, which will result in an empty set of names.
\end{enumerate*}
For the first solution \setlog returns the following:
\begin{Verbatim}
W = Y,  Names = {X,Z/N1}
Constraint: ran(B,R), subset(R,{Y}), dom(N5,N1), un(N4,N5,B),
            ran(N5,N3), subset(N3,{Y}), ran(N4,N2), Y nin N2, ...            
\end{Verbatim}
As can be seen, in this solution \verb+W = Y+ and \verb+Names+ includes \verb+X+ and \verb+Z+ plus some set \verb+N1+. As above, \verb+N1+ is the domain of \verb+N5+, a subset of \verb+B+, whose range contains only \verb+Y+. Note that the rest of \verb+B+, \verb+N4+, has a range \emph{not} containing \verb+Y+. That is, \verb+N4+ and \verb+N5+ are a partition of \verb+B+. The second solution is the following:
\begin{Verbatim}
B = {[X,Y],[Z,Y]/N2},  R = {Y/N1}
Constraint: Y neq W, dom(N5,Names), un(N5,N6,N2), ran(N2,N1), 
            subset(N1,{Y}), ran(N5,N4), subset(N4,{W}), ...
\end{Verbatim}
In this case, \verb+Y neq W+ which, along with the fact that \verb+Y+ is the only element in the range of \verb+Birthday+, implies that \verb+Names+ is the empty set. \verb+Names+ is the domain of \verb+N5+ which in turn is a subset of \verb+N2+ with a range containing only \verb+W+. Then, \verb+Names+ is necessarily empty. This is confirmed by running the following:
\begin{Verbatim}
{log}=> groundsol.
{log}=> pfun({[X,Y],[Z,Y]/B}) & ran(B,R) & subset(R,{Y}) &
        remind({[X,Y],[Z,Y]/B},W,Names) & W neq Y.
  X = n0,  Y = n2,  Z = n1,  B = {},  R = {},  W = n3,  Names = {}
\end{Verbatim}

Analysis such as these increase our confidence on the correctness of the specification perhaps more than executions where inputs are constants because variables cover all possible constant values.

\section{\label{consistency}Machine consistency}
\DefineShortVerb[commandchars=\\\$\$]{\@}

One of the advantages of having a state machine language in \setlog is that a collection of standard VC can be automatically generated.
Once discharged, these VC ensure the specification is consistent. Then, we have implemented a Verification Condition Generator (VCG) as a \setlog component. In this section we show how the VCG works.

The VCG checks several well-formedness conditions about the structure of the specification and in that case generates a file containing \setlog code encoding the VC as well as code to discharge them.
Some of the VC generated by the VCG are the following---the examples  correspond to the birthday book specification.
\begin{enumerate}
\item The initial state satisfies each and every invariant.
For instance:
\begin{Verbatim}
bbInit_sat_domBirthday :- 
  bbInit(Known,Birthday) & domBirthday(Known,Birthday).
\end{Verbatim}
\item Each operation is satisfiable and can change the state.
For example:
\begin{Verbatim}
addBirthday_is_sat :-
  addBirthday(Known,Birthday,Name,Date,Known_,Birthday_) & 
  [Known,Birthday] neq [Known_,Birthday_].
\end{Verbatim}
If the operation does not change state variables, then the VC
checks satisfiability of the operation. 
\item \emph{Invariance lemmas}: each operation preserves each and every invariant. For example\footnote{\Verb+addBirthday_pi_birthdayFun+ stands for \Verb+addBirthday+ \emph{preserves invariant} \Verb+birthdayFun+.}:
\begin{Verbatim}
addBirthday_pi_birthdayFun :-
  neg(    birthdayFun(Birthday) &
          addBirthday(Known,Birthday,Name,Date,Known_,Birthday_) 
          implies    birthdayFun(Birthday_)    ).
\end{Verbatim}
As we discuss below, some VC may be `weak' in the sense that it may be necessary to add other invariants in the premise.
\verb+addBirthday_pi_birthdayFun+ is weak.
\end{enumerate}

More VC are generated when the specification uses axioms and user-defined theorems---see \setlog user's manual for more details \cite[Section 13.6]{Rossi00}. The most important VC are the invariance lemmas.
However, if operations or invariants are unsatisfiable, then invariance lemmas will trivially hold, thus, the VCG also generates the first two VC.
Note that the first two classes of VC are satisfiability proofs whereas the third one entails unsatisfiability proofs.
In general \setlog will discharge satisfiable VC (i.e., the first two classes) in polynomial time and it will take an exponential time when discharging unsatisfiable VC (i.e., the third class), although it works well in practice as experiments and case studies suggest \citep{DBLP:journals/jar/CristiaR21,DBLP:journals/jar/CristiaR21b,DBLP:conf/birthday/CristiaR24,DBLP:journals/jar/CristiaLL23,DBLP:conf/csfw/CapozuccaCHK24}.

\subsection{Discharging verification conditions}
After loading the file generated by the VCG, users can execute a small Prolog program (called \verb+check_vcs+) that will attempt to discharge one VC after the other by passing in them to \setlog. In other words, \setlog is used as an automated theorem prover. Recall that each VC is a \setlog formula written in terms of the predicates defined by the user whose bodies are, again, \setlog formulas written in terms of constraints. Users are informed about the status of each VC:
\begin{itemize}
\item \textsc{Ok}: the VC has been successfully discharged.
\item \textsc{Timeout}: \setlog attempted to discharge the VC but it run out of time.
\item \textsc{Error}: \setlog found a counterexample contradicting the VC.
\end{itemize}

Here we analyze the second situation, whereas the third one is analyzed in Section \ref{undischarged}. \setlog can timeout when solving a VC simply because many decision procedures for set theory are NP-complete, although in practice they work in many cases\footnote{In \setlog most decision procedures are based on set unification which has been proved to be NP-complete \citep{Dovier2006}.}. When a VC times out, we do not know whether the VC is indeed a theorem of the specification or not. Then, we should help \setlog to assist us to find out whether the VC is provable or not. To this end, users can rerun the VC by calling \verb+check_vcs+ with a couple of arguments. These arguments are passed in to \setlog making it to slightly change the solving algorithm as follows.
\begin{itemize}
\item \emph{Timeout}. We can simply extend the default timeout to see if \setlog needs some more time to discharge the VC.
\item \emph{Execution options}. This is the more promising course of action as execution options influence the constraint solving algorithm. Influence means that the algorithm can take shorter to solve a formula but it also can take longer. It is hard to predict what execution options will have a positive impact in solving a particular formula. However there are some guidelines that in general produce good results. Here we will show a few execution options---see \setlog user's manual for more details \cite[Section 11.1]{Rossi00}.
\begin{itemize}
\item @subset_unify@. Implements set equality as a double set inclusion instead of implementing it by exploiting set unification \citep{Dovier2006}.
Note that set equality is pervasive in \setlog formulas even if the formula does not use it explicitly.
\item @un_fe@. Implements $un(A,B,C)$ in terms of RUQ (Section \ref{rq}). In this way \setlog produces two solutions instead of six when solving $un(A,B,\{X/C\})$, with $A$ and $B$ variables. These two solutions encode the standard proof of $x \in A \cup B$. Union is not as pervasive as equality but is one the most used constraints during the low-level stages of the algorithm.
\item @noirules@. By default \setlog applies inference rules while the goal is processed. @noirules@ deactivates these inference rules because in some cases their application may slow down the constraint solving process.
\item @strategy(ordered)@. Changes the order in which atoms are processed. It has shown to impact proofs of the form $p \land (q \lor r) \land s$. Note that invariance lemmas have this form, indeed: $\lnot(Inv \land Op \implies Inv') \equiv Inv \land Op \land \lnot Inv'$, where $Op$ is usually a disjunction---e.g., \verb+addBirthday+.
\item @tryp(prover_all)@. This is the most powerful option. It attempts to solve the VC by running it in multiple (operating system) threads. Each thread runs the VC under some combination of execution options. As soon as one thread terminates, the whole computation terminates as well. In this way, the net execution time will tend to be the time needed by the thread running the best combination of execution options for that VC.
\end{itemize}
Therefore, users can analyze the VC to determine what execution options would be useful for that particular formula. If they have access to a large computer @tryp(prover_all)@ is the best course of action as it tries all possible combinations at once. According to our experiments there is always a combination of execution options that discharges the VC in a reasonable time. At the same time, exploring execution options is far less cumbersome than attempting an interactive proof.
\end{itemize}

\subsection{\label{undischarged}Analyzing undischarged verification conditions}
\setlog may not be able to discharge a VC for a few reasons:
\begin{enumerate*}[label=(\emph{\roman*})]
\item If a VC falls outside the decision procedures implemented in \setlog, then the tool will be unable to discharge it; there is nothing to do in this case---see Section \ref{what}.
\item If the specification is wrong (e.g., a precondition is missing, an invariant is too strong, etc.) then it may be impossible to prove some VC.
\item Finally, a verification condition may require more hypotheses.
\end{enumerate*}
Let's analyze these last two situations.

Every time the status of a VC is \textsc{Error}, \setlog saves a counterexample contradicting the VC. These counterexamples can be very helpful in finding out why the proof failed. There are two kinds of counterexamples: \emph{abstract}, which may include free variables; and \emph{ground}, which do not contain free variables. For example, when \setlog attempts to discharge the VC of the birthday book specification, the VC named \verb+addBirthday_pi_birthdayFun+ fails. We can see the ground counterexample as follows\footnote{\Verb+vcgce+ stands for \emph{verification condition ground counter example}.}:
\begin{Verbatim}
{log}=> vcgce(addBirthday_pi_birthdayFun).
  Birthday = {[n2,n1]}, Known = {}, Name = n2, Date = n0,
  Known_ = {n2}, Birthday_ = {[n2,n0],[n2,n1]}
\end{Verbatim}
The reason for the failure is in this counterexample. As can be seen, \verb+Birthday =+ \verb+{[n2,n1]}+ whereas \verb+Known = {}+ thus contradicting invariant \verb+domBirthday+ which states \verb+dom(Birthday,Known)+.
That is, the counterexample says that \verb+Birthday+ is not empty while \verb+Known+, its domain according to \verb+domBirthday+, is empty. Therefore, the proof failed because it lacks \verb+domBirthday+ as an hypothesis. Once the user adds \verb+domBirthday+ as an hypothesis to \verb+addBirthday_pi_birthdayFun+ \setlog proves the VC immediately.
Although in this case the cause of the problem is a missing precondition, the same analysis is conducted when the cause ends up being an error in the specification.

This example leads us to explain an important design decision concerning the way the VCG generates invariance lemmas. Let $\{I_k\}_{k \in A}$ be the invariants of some specification and let $O$ be an operation. The invariance lemma for $O$ and $I_{k_0}$ (for some $k_0 \in A$) is\footnote{If the specification includes axioms the statement of the invariance lemma is a little bit more complex.}: $(\bigwedge_{k \in A} I_k) \land O \implies I_{k_0}'$, where $I_{k_0}' \defs I_{k_0}[\forall v \in \mathit{StateVars}: v'/v]$. However, the VCG generates the following formula: $ I_{k_0} \land O \implies I_{k_0}'$. Clearly, the invariance lemmas generated by the VCG may lack some other invariants as hypotheses. This is a design decision based on the fact that unnecessary hypotheses may severely reduce the performance of an automated theorem prover. Indeed, \setlog might start rewriting some useless hypotheses rather than the useful ones, thus spending time in dead-end proof paths. Hence, we decided that users can use counterexamples to add hypotheses one by one, if necessary.
Furthermore, \setlog provides the \verb+findh+ command family that automates the search for hypotheses. \verb+findh+ uses the abstract counterexamples saved by the VCG to find invariants that contradict them. These invariants are good candidates to become hypotheses as their presence surely will not generate the same counterexamples. As with execution options, analyzing counterexamples is far less involved than attempting interactive proofs.
It is important to observe that if a VC times out it leaves users without any clue about the reason for that.
Instead, a failed proof accompanied by a counterexample let users to work it out.

Aligned with the way \setlog generates VC, is the decision of splitting the invariant of the birthday book specification into two predicates (@domBirthday@ and @birthdayFun@). Indeed, these predicates could be defined as a single invariant. However, this strategy may hinder the automated proof process as it is like having all the invariants as hypotheses.

We have applied this methodology, for example, to analyze Chinese wall security properties \citep{DBLP:conf/csfw/CapozuccaCHK24}, where we have to add 12 hypotheses in 5 VC out of 35, that is, 30 VC need no human intervention; and to verify a model of a landing gear system \citep{DBLP:conf/birthday/CristiaR24}, where 60\% of the VC where discharged in the first run whereas the vast majority of the remaining
proofs needed only one hypothesis.

In summary, \setlog is used to automatically discharge VC and to help users to find why a VC could not be discharged.

\subsection{\label{what}What verification conditions can be discharged?}
In the previous section we mentioned that \setlog will be unable to discharge a VC if it falls outside the decision procedures implemented in it. A formal and precise description of the boundaries of the decidable fragments available in \setlog has been given in our previous work---see at the beginning of Section \ref{setlog}. In this section we give an informal account of the decision procedures implemented in \setlog. Figure \ref{fig:stack} helps us in this task. If theory \textsf{A} is above theory \textsf{B} it means that \textsf{A} inherits the signature of \textsf{B}.
\begin{itemize}
\item \textsf{LIA} stands for \emph{linear integer arithmetic}. \setlog uses an external implementation of a decision procedure for \textsf{LIA}.
\item \textsf{SET} is the Boolean algebra of sets. That is, a theory including hereditarily finite sets, equality, union, intersection, difference, set membership and their negations \citep{Dovier00}.
\item \textsf{CARD} is \textsf{SET} extended with cardinality and combined with \textsf{LIA} \citep{DBLP:journals/tplp/CristiaR23}. That is, variables denoting set cardinality can only participate in linear integer constraints. \textsf{CARD} cannot be combined with \textsf{RIS} or \textsf{RA}.
\item \textsf{FII} is \textsf{CARD} extended with finite integer intervals \citep{DBLP:journals/tocl/CristiaR24}. As with \textsf{CARD} interval limits can only participate in linear integer constraints.
\item \textsf{RIS} is \textsf{SET} extended with RIS---Section \ref{setterms}. Considering RIS of the form \verb+ris(X in A,+$\phi$\verb+)+, the theory is decidable as long as $\phi$ belongs to a decidable theory supported by \setlog. When more complex RIS are used, the condition for decidability is more involved \citep{DBLP:journals/jar/CristiaR21a}.
\item \textsf{RA} (\emph{relation algebra}) is \textsf{SET} extended with composition, converse and the identity relation. The condition for decidability is rather involved \citep{DBLP:journals/jar/CristiaR20}; we give a simplified version. \textsf{RA} is decidable for the class of formulas \emph{not} containing constraints of the form \verb+comp(R,S,R)+ or \verb+comp(S,R,R)+. As many relational constraints are defined in terms of \verb+comp+ (e.g., relational image), a formula not containing \verb+comp+ may still belong to the undecidable fragment---for example, some formulas containing (simultaneously) domain and range fall outside of the decidable fragment \citep{DBLP:journals/tcs/CantoneL14}.
\end{itemize}

\begin{figure}
\begin{center}
\begin{tikzpicture}
  [every node/.style={transform shape},font=\sffamily,
   rectangle,text centered,minimum width=2cm,minimum height=1cm]
\draw
 (0,0) node[draw,fill=gray!20] (ilp) {LIA}
 (4,0) node [draw,minimum width=6cm,fill=gray!20] (set) {SET}
 (1,1) node [draw,minimum width=4cm,fill=gray!20] (card) {CARD}
 (4.5,1) node [minimum width=1cm,fill=gray!20] {}
 (4,1) node [draw] (ris) {RIS}
 (5.5,1) node [minimum width=1cm,fill=gray!20] {}
 (6,1) node [draw] (ra) {RA}
 (1,2) node [draw,minimum width=4cm,fill=gray!20] (int) {FII}
 (-2.5,0.4) node[fill=gray!20,minimum height=10pt] (dec) {\footnotesize decidable}
 (-2.5,1) node[minimum height=10pt,draw] (undec) {\footnotesize undecidable}
 (-2.5,1.5) node[minimum height=10pt] (undec) {\footnotesize Legend};
\end{tikzpicture}
\end{center}
\caption{\label{fig:stack}The stack of theories dealt with by \setlog}
\end{figure}

The theory of restricted quantifiers, noted \textsf{RQ}, is not depicted in Figure \ref{fig:stack}. \textsf{RQ} is parametric w.r.t. some quantifier-free, decidable theory \citep{DBLP:journals/jar/CristiaR24}---for instance \textsf{RQ} could be configured with \textsf{CARD} or \textsf{LIA} or even with the decidable fragment of \textsf{RA}. However, the parameter theory must verify the following condition: if its solver, as a result of solving a formula, returns a conjunction of atoms including set variables, this conjunction must be satisfiable by substituting set variables by the empty set. This is not the case of \textsf{CARD} and \textsf{FII}. The decidability of \textsf{RQ} is as follows.
Let $\{\dots/A\}$ be the domain of a RQ where $A$ is a variable. We say that $A$ is the \emph{domain variable} of the RQ\footnote{Clearly, if the domain of a RQ is just a variable then this is the domain variable.}. The innermost formula of a RQ is called its \emph{quantifier-free formula}. For instance, @X is V + 1@ is the quantifier-free formula of the nested RQ given in \eqref{eq:rq}.
The decidable fragment of \textsf{RQ} is characterized by formulas including @foreach@ and @exists@ with a quantifier-free formula belonging to a decidable fragment and meeting at least one of
the following conditions.
\begin{enumerate}
\item The formula contains only @exists@.
\item\label{i:foreach} The formula contains only @foreach@, and none of the domain variables are used in the quantifier-free formula.
\item The formula contains nested RQ of the following form where  @foreach@ verify \eqref{i:foreach}:
\begin{Verbatim}[commandchars=\\\{\},codes={\catcode`$=3\catcode`_=8}]
exists([$X_1$ in $A_1$,...,$X_n$ in $A_n$], 
  foreach([$Y_1$ in $B_1$,...,$Y_m$ in $B_m$], $\phi$))
\end{Verbatim}
\item The formula contains @foreach@ and @exists@ but does not belong to
the above class (i.e., some @exists@ occur after some @foreach@). In this case the condition for decidability is as follows:
\begin{enumerate*}[label=(\emph{\roman*})]
\item no domain variable of a @foreach@ is used in the quantifier-free formula; and
\item\label{e:last} no @exists@ occurring \emph{after} a @foreach@ share the same domain variable.
\end{enumerate*}
For instance, the following formula \emph{does not} verify \ref{e:last}:
\Verb[commandchars=\\\[\],codes={\catcode`\$=3\catcode`_=8}]+foreach(X in {W / A}, exists(Y in {V / A}, $\phi$))+.
\end{enumerate}

If \setlog is called on a formula not meeting the conditions for the theories depicted in Figure \ref{fig:stack} nor the conditions for \textsf{RQ}, it will most likely run forever although it may terminate in some cases. Then, in general, calling \setlog on such a formula will be rather harmless. There is, however, an exception. If the formula combines \textsf{CARD} with \textsf{RIS}, \textsf{RA} or \textsf{RQ}, then the answer can be unsound.

\section{\label{testing}Test case generation}

As we have said, the state machines specified in \setlog are considered functional prototypes. These specifications will likely be implemented in some other programming language. These implementations should be tested. Model-based testing (MBT) \citep{Utting00} indicates that implementations can be tested with test cases generated from their specifications. \setlog includes an implementation of the Test Template Framework (TTF) \citep{DBLP:journals/tse/StocksC96}, a MBT method originally developed for the Z notation. The TTF was first implemented in Java for Z specifications \citep{DBLP:journals/stvr/CristiaAFPM14} and later connected to \setlog \citep{CristiaRossiSEFM13}. In that implementation users are required to write a Z specification part of which is translated into \setlog, and the results computed by \setlog are translated back to Z. Now the TTF has been implemented as a \setlog component that generates test cases from state machines specified in \setlog. As a result we have a much simpler and cleaner tool seamlessly integrated with \setlog.

\subsection{Brief introduction to the TTF}
In the TTF, test cases are generated for each operation in a state machine. 
Test case generation proceeds by applying so-called \emph{testing tactics} to the selected operation.
Each tactic partitions the \emph{input space} of the operation.
The input space of an operation is the set given by all the possible values that its input and before-state variables can assume.
What tactics are applied depends on the logical structure and mathematical elements used in the operation, as well as on the testing goals determined by the testing team.
The current implementation of the TTF provides a total of six testing tactics.
Here we will discuss only two of them, namely \emph{disjunctive normal form} (DNF) and \emph{standard partitions} (SP).
See \setlog user's manual for more details \cite[Section 13.7]{Rossi00}.

The DNF tactic requires writing the precondition of the operation into DNF and partitioning the input space in as many sets as terms are in the DNF.
In this way, the main logical alternatives of the operation are considered.
For example, when DNF is applied to \verb+addBirthday+ it will partition the input space into a set where the new person is added to the book and another set where the person already exists and so the book remains unchanged.
Hence, there will be at least one test case testing the addition of a person and another test case testing an erroneous situation.
In \setlog DNF must be the first tactic to be applied.

SP defines partitions for key mathematical operators such as union, intersection, overriding, etc. 
Figure \ref{sp:union} shows the standard partition for union, intersection and set difference.
The partition indicates conjunctions of conditions depending on the arguments of the operator.
The rationale behind SP is that set and relational operators have non trivial implementations that deserve a thorough testing.
Currently, SP supports nine set and relational operators but can be easily extended by users.

\begin{figure}
\begin{align*}
& S = \emptyset, T = \emptyset & & 
  S \neq \emptyset, T \neq \emptyset, S \subset T \\
& S = \emptyset, T \neq \emptyset & &
  S \neq \emptyset, T \neq \emptyset, T \subset S \\
& S \neq \emptyset, T = \emptyset & &
  S \neq \emptyset, T \neq \emptyset, T = S \\
& S \neq \emptyset, T \neq \emptyset, S \cap T = \emptyset & &
  S \neq \emptyset, T \neq \emptyset, S \cap T \neq \emptyset, S \not\subseteq T, T \not\subseteq S, S \neq T
\end{align*}
\caption{\label{sp:union}Standard partition for $S \cup T$, $S \cap T$ and $S \setminus T$}
\end{figure}

As a result, DNF aims at the logical structure of the operation whereas SP aims at its mathematical elements.
In \setlog each testing tactic is applied by means of its own user command, as we will show in Section \ref{ttf-bb}.

The partition of the input space is represented as a \emph{testing tree}.
Nodes in the testing tree are called \emph{test specifications}.
Each test specification is a conjunction of \setlog constraints depending on the before-state and input variables of the operation.
This conjunction is called \emph{characteristic predicate} of the test specification.
The root of the testing tree is the input space of the operation; its characteristic predicate is $true$---meaning that the test specification is not constrained in any way.
Let $C$ and $D$ be test specifications of a given testing tree, and let $\Phi_C$ and $\Phi_D$ be their characteristic predicates.
If $D$ is a child node of $C$ then $\Phi_D = \Phi_C \land \phi$ with $\phi$ being a conjunction of at least one atomic predicate different from $true$ and not present in $\Phi_C$.
In other words, the characteristic predicate of a child node includes the characteristic predicate of its father node plus the conditions added by the testing tactic from which $D$ was created.

The testing tree is a compact representation of the status of the testing process.
For example, by looking at the testing tree one can have an idea of the number of test cases that can be generated, what tactics have been applied and where.

When two or more testing tactics are applied many test specifications may turn to be unsatisfiable.
Given that it is impossible to generate test cases from unsatisfiable test specifications, they must be \emph{pruned} from the testing tree. Detecting unsatisfiable set formulas is one of the main \setlog's capabilities. Hence, the TTF provides a command, namely \verb+prunett+, that iterates over all the leaves of a testing tree and asks \setlog whether or not they are unsatisfiable. If \setlog finds an unsatisfiable test specification, \verb+prunett+ eliminates it from the testing tree.

In the TTF a test case is an assignment (or binding) of values to the input and before-state variables.
In other words, a test case is a solution to some test specification. Again, \setlog is good at finding solutions to set formulas. 
Then, \setlog provides a command, namely \verb+gentc+, that iterates over all the leaves of a testing tree and asks \setlog for a ground solution to each of them.
Test cases are derived only from the leaves of the testing tree because these nodes include all the conditions added by each applied tactic.

\subsection{\label{ttf-bb}Applying the TTF to the birthday book specification}

We will use the birthday book specification to show how the TTF works in \setlog.
Before applying the TTF on a specification, the specification has to be typechecked and the VCG has to be called on that specification.
The TTF uses information about the state machine collected by the typechecker and the VCG.
In this example we will generate test cases for \verb+addBirthday+.
As DNF is the first tactic to be applied, we issue the following command:
\begin{Verbatim}
{log}=> applydnf(addBirthday(Name,Date)).
\end{Verbatim}
That is, the command waits for a term of the form $\mathit{operation(i_1,\dots,i_n)}$ where $i_1,\dots,i_n$ are the inputs of the operation.
Users decide what arguments are inputs and what are not.
The resulting testing tree is shown with the \verb+writett+ command:
\begin{Verbatim}
addBirthday_vis
    addBirthday_dnf_1
    addBirthday_dnf_2
\end{Verbatim}
\verb+addBirthday_vis+ is the root of the tree while \verb+addBirthday_dnf_1+ and \verb+addBirthday_dnf_2+ are its children---indentation is used to depict tree levels. 
The nodes of the testing tree can be exported to a file with \verb+exporttt+, thus obtaining the following:
\begin{Verbatim}
addBirthday_dnf_1(Name,Date,Known,Birthday) :- Name nin Known.
addBirthday_dnf_2(Name,Date,Known,Birthday) :- Name in Known.
\end{Verbatim}

It would be possible to generate test cases from this testing tree but we can also apply more tactics to one or more of its test specifications, to further partition the input space.
In this case SP is applied to the \verb+addBirthday_dnf_1+ test
specification with the following command:
\begin{Verbatim}
{log}=> applysp(addBirthday_dnf_1,un(Known,{Name},Known_)).
\end{Verbatim}
The first argument is the test specification that we want to partition and the second one is a constraint present in the operation---see \verb+addBirthdayOk+ in Section \ref{statemachines}.  Then, the above command generates the following testing tree.
\begin{Verbatim}
addBirthday_vis
    addBirthday_dnf_1
        addBirthday_sp_11
        addBirthday_sp_12
        .................   % some test specifications not shown
        addBirthday_sp_17
        addBirthday_sp_18
    addBirthday_dnf_2
\end{Verbatim}
\verb+addBirthday_dnf_1+ is partitioned into eight test specifications due to the eight conjunctions of the standard partition for set union---see Figure \ref{sp:union}.
Two sample test specifications are the following:
\begin{Verbatim}
addBirthday_sp_11(Name,Date,Known,Birthday) :-
  Name nin Known & Known = {} & {Name} = {}.
addBirthday_sp_14(Name,Date,Known,Birthday) :-
  Name nin Known & Known neq {} & {Name} neq {} & disj(Known,{Name}).
\end{Verbatim}
Note that the arguments of the partition ($S$ and $T$) are substituted by the arguments of the constraint passed in to \verb+applysp+ (i.e., \verb+{Name}+ and \verb+Known+).

Now we call \verb+prunett+ to prune possible unsatisfiable test specifications. The result is the following testing tree:
\begin{Verbatim}
addBirthday_vis
    addBirthday_dnf_1
        addBirthday_sp_12
        addBirthday_sp_14
    addBirthday_dnf_2
\end{Verbatim}

Although more tactics could be applied we stop here to keep the presentation concise.
The final step in the TTF is to call \verb+gentc+ to generate one test case for each leaf in the testing tree.
The resulting test cases that can be printed with \verb+writetc+:
\begin{Verbatim}
addBirthday_tc_12(Name,Date,Known,Birthday) :-
  Name = name:n0 & Known = {}.
addBirthday_tc_14(Name,Date,Known,Birthday) :-
  Name = name:n0 & Known = {name:n1}.
addBirthday_tc_2(Name,Date,Known,Birthday) :-
  Name = name:n0 & Known = {name:n0}.
\end{Verbatim}
Observe that \verb+Birthday+ has not been bound to some value. This is so because no test specification includes a condition on that variable. Concerning these test cases, \verb+Birthday+ can assume any value. Applying SP on the postcondition \verb+un(Birthday,{[Name,Date]},Birthday_)+ as a third tactic would generate test cases with different values for \verb+Birthday+ combined with the values of these test cases. \ref{bb-ttf} presents the application of this tactic and the test cases so generated.

In summary, the TTF component uses \setlog as a back-end solver for automating two crucial but annoying, error-prone tasks: pruning unsatisfiable test  specifications and test case generation.
Users are left with the more creative task of deciding what tactics will generate the best test suite.
By implementing the TTF as a \setlog component users work with only one notation and tool.
Besides, it avoids costly translations between \setlog and Z which also brings in more efficiency.

\section{\label{conclusions}Concluding remarks}

After adding a few components we were able to turn the CLP language and solver \setlog into a formal verification tool. It seamlessly integrates implementations of some sought-after theoretical results such as symbolic execution, automated proofs and the formula-program duality. From the practical side, \setlog now offers a state machine specification language based on FOL and set theory, an environment that simplifies the analysis of functional scenarios, a verification condition generator that fed proof obligations into \setlog, and a model-based test case generator that uses \setlog as the back-end solver. There is just a single language for programs, specifications and properties; the same solver runs functional scenarios, proves properties of the specification and generates test cases.

Nonetheless, \setlog needs more functions to get closer to tools such as \agda and \dafny. To name some of them:
\begin{enumerate*}
\item Now users have to manually add hypotheses to VC which is cumbersome and error-prone. A user command could perform safety checks before the hypothesis is added.
\item Currently, \setlog does not support refinement proofs, at least not in a structured fashion.
\end{enumerate*}

More advanced features should also be analyzed. For instance, first-order automated reasoning, such as the one performed by the E-prover \citep{Schulz2019}, may help in finding missing hypotheses.
That is, \setlog could rest on E-prover to select a smaller set of promising hypotheses and then try to discharge the VC using its own engine.
An even more challenging issue is to address the proofs of VC falling outside of the decision procedures available in \setlog. An obvious solution would be to rest on an external interactive theorem prover (ITP). But using an external ITP brings in the problem of translating proof terms back and forth, thus introducing new sources of errors. On the other hand, even such proofs will have subgoals fitting in the decidable fragments where \setlog can be handy. Therefore, we have developed a simple prototype ITP on top of \setlog where users can call it whenever they have a subgoal fitting in some decidable fragment \citep{DBLP:journals/cj/CristiaKR22}. Nevertheless, this idea needs considerably more work.

\vspace{1cm}
\noindent \emph{Competing interests: The authors declare none}

\bibliographystyle{tlplike}
\bibliography{/home/mcristia/escritos/biblio}

\appendix

\section{\label{bb}One more operation of the birthday book}

The remaining operation of the birthday book shows the birthday of a given person.
\begin{Verbatim}
findBirthdayOk(Known,Birthday,Name,Date) :- 
  Name in Known & applyTo(Birthday,Name,Date).
\end{Verbatim}
\begin{Verbatim}
notAFriend(Known,Name) :- Name nin Known.
\end{Verbatim}
\begin{Verbatim}
operation(findBirthday).
findBirthday(Known,Birthday,Name,Date) :-
  findBirthdayOk(Known,Birthday,Name,Date) or notAFriend(Known,Name).
\end{Verbatim}

\section[fragile]{\label{bb-ttf}Extended application of the TTF to the \Verb+addBirthday+ operation}
The following TTF script has been applied to the \Verb+addBirthday+ operation of the birthday book specification.
\begin{verbatim}
applydnf(addBirthday(Name,Date)).
applysp(addBirthday_dnf_1,un(Known,{Name},Known_)).
prunett.
applysp(addBirthday_dnf_1,un(Birthday,{[Name,Date]},Birthday_)).
prunett.
gentc.
\end{verbatim}

In this way the following nine test cases are generated.
\begin{verbatim}
addBirthday_tc_122(Name,Date,Known,Birthday) :-
  Name=name:n1 & Known={} & Birthday={} & Date=date:n0.

addBirthday_tc_124(Name,Date,Known,Birthday) :-
  Name=name:n1 & Known={} & Birthday={[name:n0,date:n0]} & Date=date:n2.

addBirthday_tc_126(Name,Date,Known,Birthday) :-
  Name=name:n1 & Known={} &
  Birthday={[name:n1,date:n2],[name:n0,date:n0]} &
  Date=date:n2.

addBirthday_tc_127(Name,Date,Known,Birthday) :-
  Name=name:n1 & Known={} & Birthday={[name:n1,date:n0]} & Date=date:n0.

addBirthday_tc_142(Name,Date,Known,Birthday) :-
  Name=name:n0 & Known={name:n1} & Birthday={} & Date=date:n2.

addBirthday_tc_144(Name,Date,Known,Birthday) :-
  Name=name:n0 & Known={name:n1} & 
  Birthday={[name:n2,date:n2]} &
  Date=date:n3.

addBirthday_tc_146(Name,Date,Known,Birthday) :-
  Name=name:n0 & Known={name:n1} &
  Birthday={[name:n0,date:n3],[name:n2,date:n2]} &
  Date=date:n3.

addBirthday_tc_147(Name,Date,Known,Birthday) :-
  Name=name:n0 & Known={name:n1} &
  Birthday={[name:n0,date:n2]} &
  Date=date:n2.

addBirthday_tc_2(Name,Date,Known,Birthday) :-
  Name=name:n0 & Known={name:n0}.
\end{verbatim}

\end{document}